\documentclass[journal]{IEEEtran}
\usepackage[cmex10]{amsmath}
\usepackage{amsthm}
\usepackage{amssymb}
\usepackage{array}
\usepackage{graphicx}
\usepackage{cite}
\usepackage{hyperref}
\newtheorem{thm}{Theorem}[section]
\newtheorem{cor}[thm]{Corollary}
\newtheorem{lem}[thm]{Lemma}
\newtheorem{prop}[thm]{Proposition}
\theoremstyle{definition}
\newtheorem{defn}[thm]{Definition}
\theoremstyle{remark}
\newtheorem{rem}[thm]{Remark}
\numberwithin{equation}{section}

\begin{document}
\title{Passive states optimize the output of bosonic Gaussian quantum channels}
\author{Giacomo De Palma, Dario Trevisan, Vittorio Giovannetti
\thanks{G. De Palma is with NEST, Scuola Normale Superiore, Istituto Nanoscienze-CNR and INFN, I-56126 Pisa, Italy.}
\thanks{D. Trevisan is with Universit\`a degli Studi di Pisa, I-56126 Pisa, Italy.}
\thanks{V. Giovannetti is with NEST, Scuola Normale Superiore and Istituto Nanoscienze-CNR, I-56126 Pisa, Italy.}}

\maketitle
\begin{abstract}
An ordering between the quantum states emerging from a single mode gauge-covariant bosonic Gaussian channel is proven. Specifically, we show that within the set of input density matrices with the same given spectrum, the element {\it passive} with respect to the Fock basis (i.e. diagonal with decreasing eigenvalues) produces an output which majorizes all the other outputs emerging from the same set. When applied to pure input states, our finding includes as a special case the result of A. Mari, {\it et al.}, Nat. Comm. {\bf 5}, 3826 (2014) which implies that the output associated to  the vacuum majorizes the others.
\end{abstract}

\section{Introduction}
The minimum von Neumann entropy at the output of a quantum communication channel can be crucial for the determination of its classical communication capacity \cite{holevo2013quantum}.

Most communication schemes encode the information into pulses of electromagnetic radiation, that travels through metal wires, optical fibers or free space and is unavoidably affected by attenuation and noise.
Gauge-covariant quantum Gaussian channels \cite{holevo2013quantum} provide a faithful model for these effects, and are characterized by the property of preserving the thermal states of electromagnetic radiation.

It has been recently proved \cite{mari2014quantum,giovannetti2015solution,giovannetti2015majorization,holevo2015gaussian} that the output entropy of any gauge-covariant Gaussian quantum channel is minimized when the input state is the vacuum.
This result has permitted the determination of the classical information capacity of this class of channels \cite{giovannetti2014ultimate}.
However, it is not sufficient to determine the capacity region of the quantum Gaussian broadcast channel \cite{guha2007classicalproc,guha2007classical} and the triple trade-off region of the quantum-limited Gaussian attenuator \cite{wilde2012quantum,wilde2012information}.
Indeed, solving these problems would require to prove that Gaussian thermal input states minimize the output von Neumann entropy of a quantum-limited Gaussian attenuator among all the states with a given entropy.
This still unproven result would follow from a stronger conjecture, the Entropy Photon-number Inequality (EPnI) \cite{guha2008capacity,guha2008entropy}, stating that Gaussian states minimize the output von Neumann entropy of a beamsplitter among all the couples of input states, each one with a given entropy.
So far, it has been possible to prove only a weaker version of the EPnI, the quantum Entropy Power Inequality \cite{konig2014entropy,konig2013limits,de2014generalization,de2015multimode}, that provides a lower bound to the output entropy of a beamsplitter, but is never saturated.

Actually, Ref.'s \cite{mari2014quantum,giovannetti2015majorization,holevo2015gaussian} do not only prove that the vacuum minimizes the output entropy of any gauge-covariant quantum Gaussian channel.
They also prove that the output generated by the vacuum majorizes the output generated by any other state, i.e. applying a convex combination of unitary operators to the former, we can obtain any of the latter states.
This paper goes in the same direction, and proves a generalization of this result valid for any one-mode gauge-covariant quantum Gaussian channel.
Our result states that the output generated by any quantum state is majorized by the output generated by the state with the same spectrum diagonal in the Fock basis and with decreasing eigenvalues, i.e. by the state which is {\it passive} \cite{pusz1978passive,lenard1978thermodynamical,gorecki1980passive} with respect to the number operator (see \cite{vinjanampathy2015quantum,goold2015role,binder2015quantum} for the use of passive states in the context of quantum thermodynamics).
This can be understood as follows: among all the states with a given spectrum, the one diagonal in the Fock basis with decreasing eigenvalues produces the less noisy output.
Since all the states with the same spectrum have the same von Neumann entropy, our result implies that the input state with given entropy minimizing the output entropy is certainly diagonal in the Fock basis, and then reduces the minimum output entropy quantum problem to a problem on discrete classical probability distributions.

Thanks to the classification of one-mode Gaussian channels in terms of unitary equivalence \cite{holevo2013quantum}, we extend the result to the channels that are not gauge-covariant with the exception of the singular cases $A_2)$ and $B_1)$, for which we show that an optimal basis does not exist.

We also point out that the classical channel acting on discrete probability distributions associated to the restriction of the quantum-limited attenuator to states diagonal in the Fock basis coincides with the channel already known in the probability literature under the name of thinning.
First introduced by R{\'e}nyi \cite{renyi1956characterization} as a discrete analogue of the rescaling of a continuous random variable, the thinning has been recently involved in discrete versions of the central limit theorem \cite{harremoes2007thinning,yu2009monotonic,harremoes2010thinning}
and of the Entropy Power Inequality \cite{yu2009concavity,johnson2010monotonicity}.
In particular, the Restricted Thinned Entropy Power Inequality \cite{johnson2010monotonicity} states that the Poissonian probability distribution minimizes the output Shannon entropy of the thinning among all the ultra log-concave input probability distributions with a given Shannon entropy.
The techniques of this proof could be useful to prove that Gaussian thermal states minimize the output von Neumann entropy of a quantum-limited attenuator among all the input states diagonal in the Fock basis with a given von Neumann entropy (but without the ultra log-concavity constraint).
Then, thanks to the main result of this paper it would automatically follow that Gaussian thermal states minimize the output von Neumann entropy of a quantum-limited attenuator among all the input states with a given von Neumann entropy, not necessarily diagonal in the Fock basis.

The paper is organized as follows.
In Section \ref{defs} we introduce the Gaussian quantum channels, and in Section \ref{secmaj} the majorization.
The Fock rearrangement is defined in Section \ref{rearrangement}, while Section \ref{secoptimal} defines the notion of Fock optimality and proves some of its properties.
The main theorem is proved in Section \ref{mainproof}, and the case of a generic not gauge-covariant Gaussian channel is treated in Section \ref{generic}.
Finally, Section \ref{secthinning} links our result to the thinning operation.

\section{Basic definitions}\label{defs}
In this section we recall some basic definitions and facts on Gaussian quantum channels.
The interested reader can find more details in the books \cite{holevo2013quantum,barnett2002methods,holevo2011probabilistic}.
\begin{defn}[Trace norm]
The trace norm of an operator $\hat{X}$ is
\begin{equation}
\left\|\hat{X}\right\|_1:=\mathrm{Tr}\sqrt{\hat{X}^\dag\hat{X}}\;.
\end{equation}
If $\left\|\hat{X}\right\|_1$ is finite, we say that $\hat{X}$ is a trace-class operator.
\end{defn}
\begin{defn}[Quantum operation]
A quantum operation is a linear completely positive map on trace-class operators continuous in the trace norm.
\end{defn}
\begin{rem}
A trace-preserving quantum operation is a quantum channel.
\end{rem}
We consider the Hilbert space $\mathcal{H}$ of the  harmonic oscillator, i.e. the irreducible representation of the canonical commutation relation
\begin{equation}\label{CCR}
\left[\hat{a},\;\hat{a}^\dag\right]=\hat{\mathbb{I}}\;.
\end{equation}
$\mathcal{H}$ has a countable orthonormal basis
\begin{equation}
\{|n\rangle\}_{n\in\mathbb{N}}\;,\qquad \langle m|n\rangle=\delta_{mn}
\end{equation}
called the Fock basis, on which the ladder operators act as
\begin{subequations}
\begin{eqnarray}\label{acta}
\hat{a}\;|n\rangle &=& \sqrt{n}\;|n-1\rangle\\
\hat{a}^\dag\;|n\rangle &=& \sqrt{n+1}\;|n+1\rangle\;.\label{actadag}
\end{eqnarray}
\end{subequations}
We can define a number operator
\begin{equation}
\hat{N}=\hat{a}^\dag\hat{a}\;,
\end{equation}
satisfying
\begin{equation}
\hat{N}\;|n\rangle=n\;|n\rangle\;.
\end{equation}

\begin{defn}[Hilbert-Schmidt norm]
The Hilbert-Schmidt norm of an operator $\hat{X}$ is
\begin{equation}
\left\|\hat{X}\right\|_2^2:=\mathrm{Tr}\left[\hat{X}^\dag\;\hat{X}\right]\;.
\end{equation}
\end{defn}
\begin{defn}(Hilbert-Schmidt dual)\label{dagdef}
Let $\Phi$ be a linear map acting on trace class operators and continuous in the trace norm.
Its Hilbert-Schmidt dual $\Phi^\dag$ is the map on bounded operators continuous in the operator norm defined by
\begin{equation}
\mathrm{Tr}\left[\hat{Y}\;\Phi\left(\hat{X}\right)\right]=\mathrm{Tr}\left[\Phi^\dag\left(\hat{Y}\right)\;\hat{X}\right]
\end{equation}
for any trace-class operator $\hat{X}$ and any bounded operator $\hat{Y}$.
\end{defn}
\begin{defn}[Characteristic function]
The characteristic function of a trace-class operator $\hat{X}$ is
\begin{equation}
\chi_{\hat{X}}(z):=\mathrm{Tr}\left[e^{z\;\hat{a}^\dag-\bar{z}\;\hat{a}}\;\hat{X}\right]\;,\qquad z\in\mathbb{C}\;,
\end{equation}
where $\bar{z}$ denotes the complex conjugate.
\end{defn}
It is possible to prove that any trace-class operator is completely determined by its characteristic function.
\begin{thm}[Noncommutative Parceval relation]
The characteristic function provides an isometry between the Hilbert-Schmidt product and the scalar product in $L^2(\mathbb{C})$, i.e. for any two trace-class operators $\hat{X}$ and $\hat{Y}$,
\begin{equation}\label{Trint}
\mathrm{Tr}\left[\hat{X}^\dag\;\hat{Y}\right]=\int_{\mathbb{C}}\overline{\chi_{\hat{X}}(z)}\;\chi_{\hat{Y}}(z)\;\frac{d^2z}{\pi}\;.
\end{equation}
\begin{proof}
See e.g. Theorem 5.3.3 of \cite{holevo2011probabilistic}.
\end{proof}
\end{thm}
\begin{defn}[Gaussian gauge-covariant quantum channel]\label{gaugecovch}
A gauge-covariant quantum Gaussian channel with parameters $\lambda\geq0$ and $N\geq0$ can be defined by its action on the characteristic function: for any trace-class operator $\hat{X}$,
\begin{equation}\label{chiPhi}
\chi_{\Phi\left(\hat{X}\right)}(z)=e^{-|\lambda-1|\left(N+\frac{1}{2}\right)|z|^2}\;\chi_{\hat{X}}\left(\sqrt{\lambda}\;z\right)\;.
\end{equation}
The channel is called quantum-limited if $N=0$.
If $0\leq\lambda\leq1$, it is a quantum-limited attenuator, while if $\lambda\geq1$ it is a quantum-limited amplifier.
\end{defn}
\begin{lem}
Any gauge-covariant quantum Gaussian channel is continuous also in the Hilbert-Schmidt norm.
\begin{proof}
Easily follows from its action on the characteristic function \eqref{chiPhi} and the isometry \eqref{Trint}.
\end{proof}
\end{lem}
\begin{lem}\label{comp}
Any  gauge-covariant quantum Gaussian channel can be written as a quantum-limited amplifier composed with a quantum-limited attenuator.
\begin{proof}
See \cite{garcia2012majorization,giovannetti2015solution,mari2014quantum,holevo2015gaussian}.
\end{proof}
\end{lem}
\begin{lem}\label{attdag}
The Hilbert-Schmidt dual of the  quantum-limited attenuator of parameter $0<\lambda\leq1$ is $1/\lambda$ times the  quantum-limited amplifier of parameter $\lambda'=1/\lambda\geq1$, hence its restriction to trace-class operators is continuous in the trace-norm.
\begin{proof}
Easily follows from the action of the quantum-limited attenuator and amplifier on the characteristic function \eqref{chiPhi} and formula \eqref{Trint}; see also \cite{ivan2011operator}.
\end{proof}
\end{lem}
\begin{lem}
The quantum-limited attenuator of parameter $0\leq\lambda\leq1$ admits the explicit representation
\begin{equation}\label{kraus}
\Phi_\lambda\left(\hat{X}\right)=\sum_{l=0}^\infty\frac{(1-\lambda)^l}{l!}\;\lambda^\frac{\hat{N}}{2}\;\hat{a}^l\;\hat{X}\;\left(\hat{a}^\dag\right)^l\;\lambda^\frac{\hat{N}}{2}
\end{equation}
for any trace-class operator $\hat{X}$.
Then, if $\hat{X}$ is diagonal in the Fock basis, $\Phi_\lambda\left(\hat{X}\right)$ is diagonal in the same basis for any $0\leq\lambda\leq1$ also.
\begin{proof}
The channel $\Phi_\lambda$ admits the Kraus decomposition (see Eq. (4.5) of \cite{ivan2011operator})
\begin{equation}
\Phi_\lambda\left(\hat{X}\right)=\sum_{l=0}^\infty\hat{B}_l\;\hat{X}\;\hat{B}_l^\dag\;,
\end{equation}
where
\begin{equation}
\hat{B}_l=\sum_{m=0}^\infty\sqrt{\binom{m+l}{l}}\;(1-\lambda)^\frac{l}{2}\;\lambda^\frac{m}{2}\;|m\rangle\langle m+l|\;,\qquad l\in\mathbb{N}\;.
\end{equation}
Using \eqref{acta}, we have
\begin{equation}
\hat{a}^l=\sum_{m=0}^\infty\sqrt{l!\;\binom{m+l}{l}}\;|m\rangle\langle m+l|\;,
\end{equation}
and the claim easily follows.
\end{proof}
\end{lem}
\begin{lem}
The quantum-limited attenuator of parameter $\lambda=e^{-t}$ with $t\geq0$ can be written as the exponential of a Lindbladian $\mathcal{L}$, i.e. $\Phi_\lambda=e^{t\mathcal{L}}$, where
\begin{equation}\label{lindblad}
\mathcal{L}\left(\hat{X}\right)=\hat{a}\;\hat{X}\;\hat{a}^\dag-\frac{1}{2}\hat{a}^\dag\hat{a}\;\hat{X}-\frac{1}{2}\hat{X}\;\hat{a}^\dag\hat{a}
\end{equation}
for any trace-class operator $\hat{X}$.
\begin{proof}
Putting $\lambda=e^{-t}$ into \eqref{kraus} and differentiating with respect to $t$ we have for any trace-class operator $\hat{X}$
\begin{equation}
\frac{d}{dt}\Phi_{\lambda}\left(\hat{X}\right)=\mathcal{L}\left(\Phi_{\lambda}\left(\hat{X}\right)\right)\;,
\end{equation}
where $\mathcal{L}$ is the Lindbladian given by \eqref{lindblad}.
\end{proof}
\end{lem}

\begin{lem}
Let
\begin{equation}\label{Xdiag}
\hat{X}=\sum_{k=0}^\infty x_k\;|\psi_k\rangle\langle\psi_k|\;,\quad\langle\psi_k|\psi_l\rangle=\delta_{kl}\;,\quad x_0\geq x_1\geq\ldots
\end{equation}
be a self-adjoint Hilbert-Schmidt operator.
Then, the projectors
\begin{equation}\label{Pin}
\hat{\Pi}_n=\sum_{k=0}^n |\psi_k\rangle\langle\psi_k|
\end{equation}
satisfy
\begin{equation}
\mathrm{Tr}\left[\hat{\Pi}_n\;\hat{X}\right]=\sum_{k=0}^n x_k\;.
\end{equation}
\begin{proof}
Easily follows from an explicit computation.
\end{proof}
\end{lem}
\begin{lem}[Ky Fan's Maximum Principle]\label{sumeig}
Let $\hat{X}$ be a positive Hilbert-Schmidt operator with eigenvalues $\{x_k\}_{k\in\mathbb{N}}$ in decreasing order, i.e. $x_0\geq x_1\geq\ldots\;$,
and let $\hat{P}$ be a projector of rank $n+1$.
Then
\begin{equation}\label{TrPiX}
\mathrm{Tr}\left[\hat{P}\;\hat{X}\right]\leq\sum_{k=0}^n x_k\;.
\end{equation}
\begin{proof}
(See also \cite{bhatia2013matrix,fan1951maximum}).
Let us diagonalize $\hat{X}$ as in \eqref{Xdiag}.
The proof proceeds by induction on $n$.
Let $\hat{P}$ have rank one.
Since
\begin{equation}
\hat{X}\leq x_0\;\hat{\mathbb{I}}\;,
\end{equation}
we have
\begin{equation}
\mathrm{Tr}\left[\hat{P}\;\hat{X}\right]\leq x_0\;.
\end{equation}
Suppose now that \eqref{TrPiX} holds for any rank-$n$ projector.
Let $\hat{P}$ be a projector of rank $n+1$.
Its support then certainly contains a vector $|\psi\rangle$ orthogonal to the support of $\hat{\Pi}_{n-1}$, that has rank $n$.
We can choose $|\psi\rangle$ normalized (i.e. $\langle\psi|\psi\rangle=1$), and define the rank-$n$ projector
\begin{equation}
\hat{Q}=\hat{P}-|\psi\rangle\langle\psi|\;.
\end{equation}
By the induction hypothesis on $\hat{Q}$,
\begin{equation}\label{ineqQpsi}
\mathrm{Tr}\left[\hat{P}\,\hat{X}\right]=\mathrm{Tr}\left[\hat{Q}\,\hat{X}\right]+\langle\psi|\hat{X}|\psi\rangle\leq\sum_{k=0}^{n-1}x_k+\langle\psi|\hat{X}|\psi\rangle\;.
\end{equation}
Since $|\psi\rangle$ is in the support of $\hat{\mathbb{I}}-\hat{\Pi}_{n-1}$, and
\begin{equation}
\left(\hat{\mathbb{I}}-\hat{\Pi}_{n-1}\right)\hat{X}\left(\hat{\mathbb{I}}-\hat{\Pi}_{n-1}\right)\leq x_n\;\hat{\mathbb{I}}\;,
\end{equation}
we have
\begin{equation}\label{ineqpsi}
\langle\psi|\hat{X}|\psi\rangle\leq x_n\;,
\end{equation}
and this concludes the proof.
\end{proof}
\end{lem}

\begin{lem}\label{HS*}
Let $\hat{X}$ and $\hat{Y}$ be positive Hilbert-Schmidt operators with eigenvalues in decreasing order $\{x_n\}_{n\in\mathbb{N}}$ and $\{y_n\}_{n\in\mathbb{N}}$, respectively.
Then,
\begin{equation}
\sum_{n=0}^\infty (x_n-y_n)^2\leq\left\|\hat{X}-\hat{Y}\right\|_2^2\;.
\end{equation}
\begin{proof}
We have
\begin{equation}\label{TrXYr}
\left\|\hat{X}-\hat{Y}\right\|_2^2-\sum_{n=0}^\infty (x_n-y_n)^2=2\sum_{n=0}^\infty x_ny_n-2\mathrm{Tr}\left[\hat{X}\hat{Y}\right]\geq0\,.
\end{equation}
To prove the inequality in \eqref{TrXYr}, let us diagonalize $\hat{X}$ as in \eqref{Xdiag}.
We then also have
\begin{equation}
\hat{X}=\sum_{n=0}^\infty\left(x_n-x_{n+1}\right)\hat{\Pi}_n\;,
\end{equation}
where
\begin{equation}
\hat{\Pi}_n=\sum_{k=0}^n|\psi_k\rangle\langle\psi_k|\;.
\end{equation}
We then have
\begin{eqnarray}
\mathrm{Tr}\left[\hat{X}\;\hat{Y}\right] &=& \sum_{n=0}^\infty\left(x_n-x_{n+1}\right)\mathrm{Tr}\left[\hat{\Pi}_n\;\hat{Y}\right]\nonumber\\
&\leq& \sum_{n=0}^\infty\left(x_n-x_{n+1}\right)\sum_{k=0}^n y_k\nonumber\\
&=& \sum_{n=0}^\infty x_n\;y_n\;,
\end{eqnarray}
where we have used Ky Fan's Maximum Principle (Lemma \ref{sumeig}) and rearranged the sum (see also the Supplemental Material of \cite{koenig2009strong}).
\end{proof}
\end{lem}

\section{Majorization}\label{secmaj}
We recall here the definition of majorization.
The interested reader can find more details in the dedicated book \cite{marshall2010inequalities}, that however deals only with the finite-dimensional case.
\begin{defn}[Majorization]
Let $x$ and $y$ be decreasing summable sequences of positive numbers.
We say that $x$ weakly sub-majorizes $y$, or $x\succ_w y$, iff
\begin{equation}
\sum_{i=0}^n x_i\geq\sum_{i=0}^n y_i\quad\forall\;n\in\mathbb{N}\;.
\end{equation}
If they have also the same sum, we say that $x$ majorizes $y$, or $x\succ y$.
\end{defn}
\begin{defn}
Let $\hat{X}$ and $\hat{Y}$ be positive trace-class operators with eigenvalues in decreasing order $\{x_n\}_{n\in\mathbb{N}}$ and $\{y_n\}_{n\in\mathbb{N}}$, respectively.
We say that $\hat{X}$ weakly sub-majorizes $\hat{Y}$, or $\hat{X}\succ_w\hat{Y}$, iff $x\succ_w y$.
We say that $\hat{X}$ majorizes $\hat{Y}$, or $\hat{X}\succ\hat{Y}$, if they have also the same trace.
\end{defn}
From an operational point of view, majorization can also be defined with:
\begin{thm}
Given two positive operators $\hat{X}$ and $\hat{Y}$ with the same finite trace, the following conditions are equivalent:
\begin{enumerate}
  \item $\hat{X}\succ\hat{Y}$;
  \item For any continuous nonnegative convex function $f:[0,\infty)\to\mathbb{R}$ with $f(0)=0\,$,
  \begin{equation}\label{Trf}
  \mathrm{Tr}\;f\left(\hat{X}\right)\geq\mathrm{Tr}\;f\left(\hat{Y}\right)\;;
  \end{equation}
  \item For any continuous nonnegative concave function $g:[0,\infty)\to\mathbb{R}$ with $g(0)=0\,$,
  \begin{equation}\label{Trg}
  \mathrm{Tr}\;g\left(\hat{X}\right)\leq\mathrm{Tr}\;g\left(\hat{Y}\right)\;;
  \end{equation}
  \item $\hat{Y}$ can be obtained applying to $\hat{X}$ a convex combination of unitary operators, i.e. there exists a probability measure $\mu$ on unitary operators such that
  \begin{equation}
  \hat{Y}=\int\hat{U}\,\hat{X}\,\hat{U}^\dag\;d\mu\left(\hat{U}\right)\;.
  \end{equation}
\end{enumerate}
\begin{proof}
See Theorems 5, 6 and 7 of \cite{wehrl1974chaotic}.
Notice that Ref. \cite{wehrl1974chaotic} uses the opposite definition of the symbol ``$\succ$'' with respect to most literature (and to Ref. \cite{marshall2010inequalities}), i.e. there $\hat{X}\succ\hat{Y}$ means that $\hat{X}$ is majorized by $\hat{Y}$.
\end{proof}
\end{thm}
\begin{rem}
If $\hat{X}$ and $\hat{Y}$ are quantum states (i.e. their trace is one), \eqref{Trg} implies that the von Neumann entropy of $\hat{X}$ is lower than the von Neumann entropy of $\hat{Y}$, while \eqref{Trf} implies the same for all the R{\'e}nyi entropies \cite{holevo2013quantum}.
\end{rem}

\section{Fock rearrangement}\label{rearrangement}
In order to state our main theorem, we need to define
\begin{defn}[Fock rearrangement]
Let $\hat{X}$ be a positive trace-class operator with eigenvalues $\{x_n\}_{n\in\mathbb{N}}$ in decreasing order.
We define its Fock rearrangement as
\begin{equation}
\hat{X}^\downarrow:=\sum_{n=0}^\infty x_n\;|n\rangle\langle n|\;.
\end{equation}
If $\hat{X}$ coincides with its own Fock rearrangement, i.e. $\hat{X}=\hat{X}^\downarrow$, we say that it is {\it passive} \cite{pusz1978passive,lenard1978thermodynamical,gorecki1980passive} with respect to the Hamiltonian $\hat{N}$.
For simplicity, in the following we will always assume $\hat{N}$ to be the reference Hamiltonian, and an operator with $\hat{X}=\hat{X}^\downarrow$ will be called simply passive.
\end{defn}
\begin{rem}
The Fock rearrangement of any projector $\hat{\Pi}_n$ of rank $n+1$ is the projector onto the first $n+1$ Fock states:
\begin{equation}\label{Pin*}
\hat{\Pi}_n^\downarrow=\sum_{i=0}^n|i\rangle\langle i|\;.
\end{equation}
\end{rem}

We define the notion of passive-preserving quantum operation, that will be useful in the following.
\begin{defn}[Passive-preserving quantum operation]\label{*pres}
We say that a quantum operation $\Phi$ is passive-preserving if $\Phi\left(\hat{X}\right)$ is passive for any passive positive trace-class operator $\hat{X}$.
\end{defn}

We will also need these lemmata:
\begin{lem}\label{PXPlem}
For any self-adjoint trace-class operator $\hat{X}$,
\begin{equation}
\lim_{N\to\infty}\left\|\hat{\Pi}_N^\downarrow\;\hat{X}\;\hat{\Pi}_N^\downarrow-\hat{X}\right\|_2=0\;,
\end{equation}
where the $\hat{\Pi}_N^\downarrow$ are the projectors onto the first $N+1$ Fock states defined in \eqref{Pin*}.
\begin{proof}
We have
\begin{eqnarray}\label{PXP}
\left\|\hat{\Pi}_N^\downarrow\hat{X}\hat{\Pi}_N^\downarrow-\hat{X}\right\|_2^2 &=& \mathrm{Tr}\left[\hat{X}\left(\hat{\mathbb{I}}+\hat{\Pi}_N^\downarrow\right)\hat{X}\left(\hat{\mathbb{I}}-\hat{\Pi}_N^\downarrow\right)\right]\nonumber\\  &\leq&2\;\mathrm{Tr}\left[\hat{X}^2\left(\hat{\mathbb{I}}-\hat{\Pi}_N^\downarrow\right)\right]\nonumber\\
&=&2\sum_{n=N+1}^\infty\langle n|\hat{X}^2|n\rangle\;,
\end{eqnarray}
where we have used that
\begin{equation}
\hat{\mathbb{I}}+\hat{\Pi}_N^\downarrow\leq 2\;\hat{\mathbb{I}}\;.
\end{equation}
Since $\hat{X}$ is trace-class, it is also Hilbert-Schmidt, the sum in \eqref{PXP} converges, and its tail tends to zero for $N\to\infty$.
\end{proof}
\end{lem}
\begin{lem}\label{*symtr}
A positive trace-class operator $\hat{X}$ is passive iff for any finite-rank projector $\hat{P}$
\begin{equation}\label{PP*X}
\mathrm{Tr}\left[\hat{P}\;\hat{X}\right]\leq\mathrm{Tr}\left[\hat{P}^\downarrow\;\hat{X}\right]\;.
\end{equation}
\begin{proof}
First, suppose that $\hat{X}$ is passive with eigenvalues $\{x_n\}_{n\in\mathbb{N}}$ in decreasing order, and let $\hat{P}$ have rank $n+1$.
Then, by Lemma \ref{sumeig}
\begin{equation}
\mathrm{Tr}\left[\hat{P}\;\hat{X}\right]\leq\sum_{i=0}^n x_i=\mathrm{Tr}\left[\hat{P}^\downarrow\;\hat{X}\right]\;.
\end{equation}

Suppose now that \eqref{PP*X} holds for any finite-rank projector.
Let us diagonalize $\hat{X}$ as in \eqref{Xdiag}.
Putting into \eqref{PP*X} the projectors $\hat{\Pi}_n$ defined in \eqref{Pin},
\begin{equation}
\sum_{i=0}^n x_i=\mathrm{Tr}\left[\hat{\Pi}_n\;\hat{X}\right]\leq\mathrm{Tr}\left[\hat{\Pi}_n^\downarrow\;\hat{X}\right]\leq\sum_{i=0}^n x_i\;,
\end{equation}
where we have again used Lemma \ref{sumeig}.
It follows that for any $n\in\mathbb{N}$
\begin{equation}
\mathrm{Tr}\left[\hat{\Pi}_n^\downarrow\;\hat{X}\right]=\sum_{i=0}^n x_i\;,
\end{equation}
and
\begin{equation}
\langle n|\hat{X}|n\rangle=x_n\;.
\end{equation}
It is then easy to prove by induction on $n$ that
\begin{equation}
\hat{X}=\sum_{n=0}^\infty x_n\;|n\rangle\langle n|\;,
\end{equation}
i.e. $\hat{X}$ is passive.
\end{proof}
\end{lem}
\begin{lem}\label{sumstar}
Let $\left\{\hat{X}_n\right\}_{n\in\mathbb{N}}$ be a sequence of positive trace-class operators with $\hat{X}_n$ passive for any $n\in\mathbb{N}$.
Then also $\sum_{n=0}^\infty\hat{X}_n$ is passive, provided that its trace is finite.
\begin{proof}
Follows easily from the definition of Fock rearrangement.
\end{proof}
\end{lem}
\begin{lem}\label{phistar}
Let $\Phi$ be a quantum operation.
Suppose that $\Phi\left(\hat{\Pi}\right)$ is passive for any passive finite-rank projector $\hat{\Pi}$.
Then, $\Phi$ is passive-preserving.
\begin{proof}
Choose a passive operator
\begin{equation}
\hat{X}=\sum_{n=0}^\infty x_n\,|n\rangle\langle n|\;,
\end{equation}
with $\{x_n\}_{n\in\mathbb{N}}$ positive and decreasing.
We then also have
\begin{equation}
\hat{X}=\sum_{n=0}^\infty z_n\;\hat{\Pi}_n^\downarrow\;,
\end{equation}
where the $\hat{\Pi}_n^\downarrow$ are defined in \eqref{Pin*}, and
\begin{equation}
z_n=x_n-x_{n+1}\geq0\;.
\end{equation}
Since by hypothesis $\Phi\left(\hat{\Pi}_n^\downarrow\right)$ is passive for any $n\in\mathbb{N}$, according to Lemma \ref{sumstar} also
\begin{equation}
\Phi\left(\hat{X}\right)=\sum_{n=0}^\infty z_n\;\Phi\left(\hat{\Pi}_n^\downarrow\right)
\end{equation}
is passive.
\end{proof}
\end{lem}

\begin{lem}\label{majprojlem}
Let $\hat{X}$ and $\hat{Y}$ be positive trace-class operators.
\begin{enumerate}
\item Suppose that for any finite-rank projector $\hat{\Pi}$
\begin{equation}\label{majproj}
\mathrm{Tr}\left[\hat{\Pi}\,\hat{X}\right]\leq\mathrm{Tr}\left[\hat{\Pi}^\downarrow\,\hat{Y}\right]\;.
\end{equation}
Then $\hat{X}\prec_w\hat{Y}$.
\item Let $\hat{Y}$ be passive, and suppose that $\hat{X}\prec_w\hat{Y}$.
Then \eqref{majproj} holds for any finite-rank projector $\hat{\Pi}$.
\end{enumerate}
\begin{proof}
Let $\{x_n\}_{n\in\mathbb{N}}$ and $\{y_n\}_{n\in\mathbb{N}}$ be the eigenvalues in decreasing order of $\hat{X}$ and $\hat{Y}$, respectively, and let us diagonalize $\hat{X}$ as in \eqref{Xdiag}.
\begin{enumerate}
\item Suppose first that \eqref{majproj} holds for any finite-rank projector $\hat{\Pi}$.
For any $n\in\mathbb{N}$ we have
\begin{equation}\label{eqtr}
\sum_{i=0}^n x_i=\mathrm{Tr}\left[\hat{\Pi}_n\,\hat{X}\right]\leq\mathrm{Tr}\left[\hat{\Pi}_n^\downarrow\,\hat{Y}\right]\leq\sum_{i=0}^n y_i\;,
\end{equation}
where the $\hat{\Pi}_n$ are defined in \eqref{Pin} and we have used Lemma \ref{sumeig}.
Then $x\prec_w y$, and $\hat{X}\prec_w\hat{Y}$.

\item Suppose now that $\hat{X}\prec_w\hat{Y}$ and $\hat{Y}=\hat{Y}^\downarrow$.
Then, for any $n\in\mathbb{N}$ and any projector $\hat{\Pi}$ of rank $n+1$,
\begin{equation}
\mathrm{Tr}\left[\hat{\Pi}\,\hat{X}\right]\leq \sum_{i=0}^n x_i\leq\sum_{i=0}^n y_i=\mathrm{Tr}\left[\hat{\Pi}^\downarrow\,\hat{Y}\right]\;,
\end{equation}
where we have used Lemma \ref{sumeig} again.
\end{enumerate}
\end{proof}
\end{lem}
\begin{lem}\label{corTr}
Let $\hat{Y}$ and $\hat{Z}$ be positive trace-class operators with $\hat{Y}\prec_w\hat{Z}=\hat{Z}^\downarrow$.
Then, for any positive trace-class operator $\hat{X}$,
\begin{equation}
\mathrm{Tr}\left[\hat{X}\;\hat{Y}\right]\leq\mathrm{Tr}\left[\hat{X}^\downarrow\;\hat{Z}\right]\;.
\end{equation}
\begin{proof}
Let us diagonalize $\hat{X}$ as in \eqref{Xdiag}.
Then, it can be rewritten as
\begin{equation}\label{XPi}
\hat{X}=\sum_{n=0}^\infty d_n\,\hat{\Pi}_n\;,
\end{equation}
where the projectors $\hat{\Pi}_n$ are as in \eqref{Pin} and
\begin{equation}
d_n=x_n-x_{n+1}\geq0\;.
\end{equation}
The Fock rearrangement of $\hat{X}$ is
\begin{equation}\label{X*Pi}
\hat{X}^\downarrow=\sum_{n=0}^\infty d_n\,\hat{\Pi}_n^\downarrow\;.
\end{equation}
We then have from Lemma \ref{majprojlem}
\begin{eqnarray}
\mathrm{Tr}\left[\hat{X}\;\hat{Y}\right] &=& \sum_{n=0}^\infty d_n\;\mathrm{Tr}\left[\hat{\Pi}_n\;\hat{Y}\right]\leq\sum_{n=0}^\infty d_n\;\mathrm{Tr}\left[\hat{\Pi}_n^\downarrow\;\hat{Z}\right]\nonumber\\
&=&\mathrm{Tr}\left[\hat{X}^\downarrow\;\hat{Z}\right]\;.
\end{eqnarray}
\end{proof}
\end{lem}

\begin{lem}\label{majsum}
Let $\left\{\hat{X}_n\right\}_{n\in\mathbb{N}}$ and $\left\{\hat{Y}_n\right\}_{n\in\mathbb{N}}$ be two sequences of positive trace-class operators, with $\hat{Y}_n=\hat{Y}_n^\downarrow$ and $\hat{X}_n\prec_w\hat{Y}_n$ for any $n\in\mathbb{N}$.
Then
\begin{equation}
\sum_{n=0}^\infty\hat{X}_n\prec_w\sum_{n=0}^\infty\hat{Y}_n\;,
\end{equation}
provided that both sides have finite traces.
\begin{proof}
Let $\hat{P}$ be a finite-rank projector.
Since $\hat{X}_n\prec_w\hat{Y}_n$ and $Y_n=Y_n^\downarrow$, by the second part of Lemma \ref{majprojlem}
\begin{equation}
\mathrm{Tr}\left[\hat{P}\;\hat{X}_n\right]\leq\mathrm{Tr}\left[\hat{P}^\downarrow\;\hat{Y}_n\right]\qquad\forall\;n\in\mathbb{N}\;.
\end{equation}
Then,
\begin{equation}
\mathrm{Tr}\left[\hat{P}\;\sum_{n=0}^\infty\hat{X}_n\right]\leq\mathrm{Tr}\left[\hat{P}^\downarrow\;\sum_{n=0}^\infty\hat{Y}_n\right]\;,
\end{equation}
and the submajorization follows from the first part of Lemma \ref{majprojlem}.
\end{proof}
\end{lem}
\begin{lem}\label{HS**}
The Fock rearrangement is continuous in the Hilbert-Schmidt norm.
\begin{proof}
Let $\hat{X}$ and $\hat{Y}$ be trace-class operators, with eigenvalues in decreasing order $\{x_n\}_{n\in\mathbb{N}}$ and $\{y_n\}_{n\in\mathbb{N}}$, respectively.
We then have
\begin{equation}
\left\|\hat{X}^\downarrow-\hat{Y}^\downarrow\right\|_2^2=\sum_{n=0}^\infty(x_n-y_n)^2\leq\left\|\hat{X}-\hat{Y}\right\|_2^2\;,
\end{equation}
where we have used Lemma \ref{HS*}.
\end{proof}
\end{lem}

\section{Fock-optimal quantum operations}\label{secoptimal}
We will prove that any  gauge-covariant Gaussian quantum channel satisfies this property:
\begin{defn}[Fock-optimal quantum operation]
We say that a quantum operation $\Phi$ is Fock-optimal if for any positive trace-class operator $\hat{X}$
\begin{equation}\label{conjectureeq}
\Phi\left(\hat{X}\right)\prec_w\Phi\left(\hat{X}^\downarrow\right)\;,
\end{equation}
i.e. Fock-rearranging the input always makes the output less noisy, or among all the quantum states with a given spectrum, the passive one generates the least noisy output.
\end{defn}
\begin{rem}
If $\Phi$ is trace-preserving, weak sub-majorization in \eqref{conjectureeq} can be equivalently replaced by majorization.
\end{rem}
We can now state the main result of the paper:
\begin{thm}\label{maintheorem}
Any one-mode gauge-covariant Gaussian quantum channel is passive-preserving and Fock-optimal.
\begin{proof}
See Section \ref{mainproof}.
\end{proof}
\end{thm}
\begin{cor}
Any linear combination with positive coefficients of gauge-covariant quantum Gaussian channels is Fock-optimal.
\begin{proof}
Follows from Theorem \ref{maintheorem} and Lemma \ref{convexhull}.
\end{proof}
\end{cor}

In the remainder of this section, we prove some general properties of Fock-optimality that will be needed in the main proof.

\begin{lem}\label{majprojprop}
Let $\Phi$ be a passive-preserving quantum operation.
If for any finite-rank projector $\hat{P}$
\begin{equation}\label{hypproj}
\Phi\left(\hat{P}\right)\prec_w\Phi\left(\hat{P}^\downarrow\right)\;,
\end{equation}
then $\Phi$ is Fock-optimal.
\begin{proof}
Let $\hat{X}$ be a positive trace-class operator as in \eqref{XPi}, with Fock rearrangement as in \eqref{X*Pi}.
Since $\Phi$ is passive-preserving, for any $n\in\mathbb{N}$
\begin{equation}
\Phi\left(\hat{\Pi}_n\right)\prec_w\Phi\left(\hat{\Pi}_n^\downarrow\right)=\Phi\left(\hat{\Pi}_n^\downarrow\right)^\downarrow\;.
\end{equation}
Then we can apply Lemma \ref{majsum} to
\begin{equation}
\Phi\left(\hat{X}\right)=\sum_{n=0}^\infty d_n\;\Phi\left(\hat{\Pi}_n\right)\prec_w\sum_{n=0}^\infty d_n\;\Phi\left(\hat{\Pi}_n^\downarrow\right)=\Phi\left(\hat{X}^\downarrow\right)\;,
\end{equation}
and the claim follows.
\end{proof}
\end{lem}

\begin{lem}\label{conjPQprop}
A quantum operation $\Phi$ is passive-preserving and Fock-optimal iff
\begin{equation}\label{conjPQeq}
\mathrm{Tr}\left[\hat{Q}\;\Phi\left(\hat{P}\right)\right]\leq\mathrm{Tr}\left[\hat{Q}^\downarrow\;\Phi\left(\hat{P}^\downarrow\right)\right]
\end{equation}
for any two finite-rank projectors $\hat{Q}$ and $\hat{P}$.
\begin{proof}
Suppose first that $\Phi$ is passive-preserving and Fock-optimal, and let $\hat{P}$ and $\hat{Q}$ be finite-rank projectors.
Then
\begin{equation}
\Phi\left(\hat{P}\right)\prec_w\Phi\left(\hat{P}^\downarrow\right)=\Phi\left(\hat{P}^\downarrow\right)^\downarrow\;,
\end{equation}
and \eqref{conjPQeq} follows from Lemma \ref{majprojlem}.

Suppose now that \eqref{conjPQeq} holds for any finite-rank projectors $\hat{P}$ and $\hat{Q}$.
Choosing $\hat{P}$ passive, we get
\begin{equation}
\mathrm{Tr}\left[\hat{Q}\;\Phi\left(\hat{P}\right)\right]\leq\mathrm{Tr}\left[\hat{Q}^\downarrow\;\Phi\left(\hat{P}\right)\right]\;,
\end{equation}
and from Lemma \ref{*symtr} also $\Phi\left(\hat{P}\right)$ is passive, so from Lemma \ref{phistar} $\Phi$ is passive-preserving.
Choosing now a generic $\hat{P}$, by Lemma \ref{majprojlem}
\begin{equation}
\Phi\left(\hat{P}\right)\prec_w\Phi\left(\hat{P}^\downarrow\right)\;,
\end{equation}
and from Lemma \ref{majprojprop} $\Phi$ is also Fock-optimal.
\end{proof}
\end{lem}
We can now prove the two fundamental properties of Fock-optimality:
\begin{thm}\label{Phidag}
Let $\Phi$ be a quantum operation with the restriction of its Hilbert-Schmidt dual $\Phi^\dag$ to trace-class operators continuous in the trace norm.
Then, $\Phi$ is passive-preserving and Fock-optimal iff $\Phi^\dag$ is passive-preserving and Fock-optimal.
\begin{proof}
Condition \eqref{conjPQeq} can be rewritten as
\begin{equation}\label{PQdag}
\mathrm{Tr}\left[\Phi^\dag\left(\hat{Q}\right)\hat{P}\right]\leq\mathrm{Tr}\left[\Phi^\dag\left(\hat{Q}^\downarrow\right)\hat{P}^\downarrow\right]\;,
\end{equation}
and is therefore symmetric for $\Phi$ and $\Phi^\dag$.
\end{proof}
\end{thm}
\begin{thm}\label{qcirc}
Let $\Phi_1$ and $\Phi_2$ be passive-preserving and Fock-optimal quantum operations with the restriction of $\Phi_2^\dag$ to trace-class operators continuous in the trace norm.
Then, their composition $\Phi_2\circ\Phi_1$ is also passive-preserving and Fock-optimal.
\begin{proof}
Let $\hat{P}$ and $\hat{Q}$ be finite-rank projectors.
Since $\Phi_2$ is Fock-optimal and passive-preserving,
\begin{equation}
\Phi_2\left(\Phi_1\left(\hat{P}\right)\right)\prec_w\Phi_2\left(\Phi_1\left(\hat{P}\right)^\downarrow\right)=\Phi_2\left(\Phi_1\left(\hat{P}\right)^\downarrow\right)^\downarrow\;,
\end{equation}
and by Lemma \ref{majprojlem}
\begin{align}\label{eqphi12}
\mathrm{Tr}\left[\hat{Q}\;\Phi_2\left(\Phi_1\left(\hat{P}\right)\right)\right] &\leq \mathrm{Tr}\left[\hat{Q}^\downarrow\;\Phi_2\left(\Phi_1\left(\hat{P}\right)^\downarrow\right)\right]\nonumber\\
&= \mathrm{Tr}\left[\Phi_2^\dag\left(\hat{Q}^\downarrow\right)\Phi_1\left(\hat{P}\right)^\downarrow\right]\;.
\end{align}
Since $\Phi_1$ is Fock-optimal and passive-preserving,
\begin{equation}
\Phi_1\left(\hat{P}\right)^\downarrow\prec_w\Phi_1\left(\hat{P}^\downarrow\right)=\Phi_1\left(\hat{P}^\downarrow\right)^\downarrow\;.
\end{equation}
From Theorem \ref{Phidag} also $\Phi_2^\dag$ is passive-preserving, and $\Phi_2^\dag\left(\hat{Q}^\downarrow\right)$ is passive.
Lemma \ref{corTr} implies then
\begin{align}\label{eqphi3}
\mathrm{Tr}\left[\Phi_2^\dag\left(\hat{Q}^\downarrow\right)\Phi_1\left(\hat{P}\right)^\downarrow\right] &\leq \mathrm{Tr}\left[\Phi_2^\dag\left(\hat{Q}^\downarrow\right)\Phi_1\left(\hat{P}^\downarrow\right)\right]\nonumber\\
&= \mathrm{Tr}\left[\hat{Q}^\downarrow\;\Phi_2\left(\Phi_1\left(\hat{P}^\downarrow\right)\right)\right]\;,
\end{align}
and the claim follows from Lemma \ref{conjPQprop} combining \eqref{eqphi3} with \eqref{eqphi12}.
\end{proof}
\end{thm}

\begin{lem}\label{Phifinite}
Let $\Phi$ be a quantum operation continuous in the Hilbert-Schmidt norm.
Suppose that for any $N\in\mathbb{N}$ its restriction to the span of the first $N+1$ Fock states is passive-preserving and Fock-optimal, i.e. for any positive operator $\hat{X}$ supported on the span of the first $N+1$ Fock states
\begin{equation}
\Phi\left(\hat{X}\right)\prec_w\Phi\left(\hat{X}^\downarrow\right)=\Phi\left(\hat{X}^\downarrow\right)^\downarrow\;.
\end{equation}
Then, $\Phi$ is passive-preserving and Fock-optimal.
\begin{proof}
Let $\hat{P}$ and $\hat{Q}$ be two generic finite-rank projectors.
Since the restriction of $\Phi$ to the support of $\hat{\Pi}_N^\downarrow$ is Fock-optimal and passive-preserving,
\begin{align}
\Phi\left(\hat{\Pi}_N^\downarrow\;\hat{P}\;\hat{\Pi}_N^\downarrow\right) &\prec_w \Phi\left(\left(\hat{\Pi}_N^\downarrow\;\hat{P}\;\hat{\Pi}_N^\downarrow\right)^\downarrow\right)\nonumber\\ &=\left(\Phi\left(\left(\hat{\Pi}_N^\downarrow\;\hat{P}\;\hat{\Pi}_N^\downarrow\right)^\downarrow\right)\right)^\downarrow\;.
\end{align}
Then, from Lemma \ref{majprojlem}
\begin{equation}\label{TrPQN}
\mathrm{Tr}\left[\hat{Q}\;\Phi\left(\hat{\Pi}_N^\downarrow\;\hat{P}\;\hat{\Pi}_N^\downarrow\right)\right] \leq \mathrm{Tr}\left[\hat{Q}^\downarrow\;\Phi\left(\left(\hat{\Pi}_N^\downarrow\;\hat{P}\;\hat{\Pi}_N^\downarrow\right)^\downarrow\right)\right]\;.
\end{equation}
From Lemma \ref{PXPlem},
\begin{equation}
\left\|\hat{\Pi}_N^\downarrow\;\hat{P}\;\hat{\Pi}_N^\downarrow-\hat{P}\right\|_2\to0\qquad\text{for}\;N\to\infty\;,
\end{equation}
and since $\Phi$, the Fock rearrangement (see Lemma \ref{HS**}) and the Hilbert-Schmidt product are continuous in the Hilbert-Schmidt norm, we can take the limit $N\to\infty$ in \eqref{TrPQN} and get
\begin{equation}
\mathrm{Tr}\left[\hat{Q}\;\Phi\left(\hat{P}\right)\right] \leq \mathrm{Tr}\left[\hat{Q}^\downarrow\;\Phi\left(\hat{P}^\downarrow\right)\right]\;.
\end{equation}
The claim now follows from Lemma \ref{conjPQprop}.
\end{proof}
\end{lem}
\begin{lem}\label{convexhull}
Let $\Phi_1$ and $\Phi_2$ be Fock-optimal and passive-preserving quantum operations.
Then, also $\Phi_1+\Phi_2$ is Fock-optimal and passive-preserving.
\begin{proof}
Easily follows from Lemma \ref{conjPQprop}.
\end{proof}
\end{lem}

\section{Proof of the main theorem}\label{mainproof}
First, we can reduce the problem to the quantum-limited attenuator:
\begin{lem}\label{att->all}
If the  quantum-limited attenuator is passive-preserving and Fock-optimal, the property extends to any  gauge-covariant quantum Gaussian channel.
\begin{proof}
From Lemma \ref{comp}, any quantum  gauge-covariant Gaussian channel can be obtained composing a quantum-limited attenuator with a quantum-limited amplifier.
From Lemma \ref{attdag}, the Hilbert-Schmidt dual of a quantum-limited amplifier is proportional to a quantum-limited attenuator, and from Lemma \ref{Phidag} also the amplifier is passive-preserving and Fock-optimal.
Finally, the claim follows from Theorem \ref{qcirc}.
\end{proof}
\end{lem}

By Lemma \ref{Phifinite}, we can restrict to quantum states $\hat{\rho}$ supported on the span of the first $N+1$ Fock states.
Let now
\begin{equation}
\hat{\rho}(t)=e^{t\mathcal{L}}\left(\hat{\rho}\right)\;,
\end{equation}
where $\mathcal{L}$ is the generator of the quantum-limited attenuator defined in \eqref{lindblad}.
From the explicit representation \eqref{kraus}, it is easy to see that $\hat{\rho}(t)$ remains supported on the span of the first $N+1$ Fock states for any $t\geq0$.
In finite dimension, the quantum states with non-degenerate spectrum are dense in the set of all quantum states.
Besides, the spectrum is a continuous function of the operator, and any linear map is continuous.
Then, without loss of generality we can suppose that $\hat{\rho}$ has non-degenerate spectrum.
Let
\begin{equation}
p(t)=\left(p_0(t),\ldots,p_N(t)\right)
\end{equation}
be the vectors of the eigenvalues of $\hat{\rho}(t)$ in decreasing order, and let
\begin{equation}
s_n(t)=\sum_{i=0}^n p_i(t)\;,\qquad n=0,\ldots,\,N\;,
\end{equation}
their partial sums, that we similarly collect into the vector $s(t)$.
Let instead
\begin{equation}\label{pndt}
p_n^\downarrow(t)=\langle n|e^{t\mathcal{L}}\left(\hat{\rho}^\downarrow\right)|n\rangle\;,\qquad n=0,\,\ldots,\,N
\end{equation}
be the eigenvalues of $e^{t\mathcal{L}}\left(\hat{\rho}^\downarrow\right)$ (recall that it is diagonal in the Fock basis for any $t\geq0$), and
\begin{equation}
s_n^\downarrow(t)=\sum_{i=0}^n p_i^\downarrow(t)\;,\qquad n=0,\,\ldots,\,N\;,
\end{equation}
their partial sums.
Notice that $p(0)=p^\downarrow(0)$ and then $s(0)=s^\downarrow(0)$.
Combining \eqref{pndt} with the expression for the Lindbladian \eqref{lindblad}, with the help of \eqref{acta} and \eqref{actadag} it is easy to see that the eigenvalues $p_n^\downarrow(t)$ satisfy
\begin{equation}
\frac{d}{dt}p_n^\downarrow(t)=\left(n+1\right)p_{n+1}^\downarrow(t)-n\,p_n^\downarrow(t)\;,
\end{equation}
implying
\begin{equation}
\frac{d}{dt}s_n^\downarrow(t)=(n+1)\left(s^\downarrow_{n+1}(t)-s^\downarrow_n(t)\right)
\end{equation}
for their partial sums.
The proof of Theorem \ref{maintheorem} is a consequence of:
\begin{lem}\label{deg}
The spectrum of $\hat{\rho}(t)$ can be degenerate at most in isolated points.
\end{lem}
\begin{lem}\label{lemma1}
$s(t)$ is continuous in $t$, and for any $t\geq0$ such that $\hat{\rho}(t)$ has non-degenerate spectrum it satisfies
\begin{equation}\label{sdot}
\frac{d}{dt}s_n(t)\leq(n+1)(s_{n+1}(t)-s_n(t))\;,\qquad n=0,\,\ldots,\,N-1\;.
\end{equation}
\end{lem}
\begin{lem}\label{lemma2}
If $s(t)$ is continuous in $t$ and satisfies \eqref{sdot}, then
\begin{equation}
s_n(t)\leq s_n^\downarrow(t)
\end{equation}
for any $t\geq0$ and $n=0,\,\ldots,\,N$.
\end{lem}
Lemma \ref{lemma2} implies that the quantum-limited attenuator is passive-preserving.
Indeed, let us choose $\hat{\rho}$ passive.
Since $e^{t\mathcal{L}}\left(\hat{\rho}\right)$ is diagonal in the Fock basis, $s_n^\downarrow(t)$ is the sum of the eigenvalues corresponding to the first $n+1$ Fock states $|0\rangle,\;\ldots,\;|n\rangle$.
Since $s_n(t)$ is the sum of the $n+1$ greatest eigenvalues, $s_n^\downarrow(t)\leq s_n(t)$.
However, Lemma \ref{lemma2} implies $s_n(t)=s_n^\downarrow(t)$ for $n=0,\,\ldots,\,N$.
Thus $p_n(t)=p_n^\downarrow(t)$, so the operator $e^{t\mathcal{L}}\left(\hat{\rho}\right)$ is passive for any $t$, and the channel $e^{t\mathcal{L}}$ is passive-preserving.

Then from the definition of majorization and Lemma \ref{lemma2} again,
\begin{equation}
e^{t\mathcal{L}}\left(\hat{\rho}\right)\prec_w e^{t\mathcal{L}}\left(\hat{\rho}^\downarrow\right)
\end{equation}
for any $\hat{\rho}$, and the quantum-limited attenuator is also Fock-optimal.

\subsection{Proof of Lemma \ref{deg}}
The matrix elements of the operator $e^{t\mathcal{L}}\left(\hat{\rho}\right)$ are analytic functions of $t$.
The spectrum of $\hat{\rho}(t)$ is degenerate iff the function
\begin{equation}
\phi(t)=\prod_{i\neq j}\left(p_i(t)-p_j(t)\right)
\end{equation}
vanishes.
This function is a symmetric polynomial in the eigenvalues of $\hat{\rho}(t)=e^{t\mathcal{L}}\left(\hat{\rho}\right)$.
Then, for the Fundamental Theorem of Symmetric Polynomials (see e.g Theorem 3 in Chapter 7 of \cite{cox2015ideals}), $\phi(t)$ can be written as a polynomial in the elementary symmetric polynomials in the eigenvalues of $\hat{\rho}(t)$.
However, these polynomials coincide with the coefficients of the characteristic polynomial of $\hat{\rho}(t)$, that are in turn polynomials in its matrix elements.
It follows that $\phi(t)$ can be written as a polynomial in the matrix elements of the operator $\hat{\rho}(t)$.
Since each of these matrix element is an analytic function of $t$, also $\phi(t)$ is analytic.
Since by hypothesis the spectrum of $\hat{\rho}(0)$ is non-degenerate, $\phi$ cannot be identically zero, and its zeroes are isolated points.

\subsection{Proof of Lemma \ref{lemma1}}
The matrix elements of the operator $e^{t\mathcal{L}}\left(\hat{\rho}\right)$ are analytic (and hence continuous and differentiable) functions of $t$.
Then for Weyl's Perturbation Theorem $p(t)$ is continuous in $t$, and also $s(t)$ is continuous (see e.g. Corollary III.2.6 and the discussion at the beginning of Chapter VI  of \cite{bhatia2013matrix}).
Let $\hat{\rho}(t_0)$ have non-degenerate spectrum.
Then, $\hat{\rho}(t)$ has non-degenerate spectrum for any $t$ in a suitable neighbourhood of $t_0$.
In this neighbourhood, we can diagonalize $\hat{\rho}(t)$ with
\begin{equation}
\hat{\rho}(t)=\sum_{n=0}^N p_n(t) |\psi_n(t)\rangle\langle\psi_n(t)|\;,
\end{equation}
where the eigenvalues in decreasing order $p_n(t)$ are differentiable functions of $t$ (see Theorem 6.3.12 of \cite{horn2012matrix}),
and
\begin{equation}
\frac{d}{dt}p_n(t)=\langle\psi_n(t)|\mathcal{L}\left(\hat{\rho}(t)\right)|\psi_n(t)\rangle\;.
\end{equation}
We then have
\begin{equation}
\frac{d}{dt}s_n(t)=\mathrm{Tr}\left[\hat{\Pi}_n(t)\;\mathcal{L}\left(\hat{\rho}(t)\right)\right]\;,
\end{equation}
where
\begin{equation}
\hat{\Pi}_n(t)=\sum_{i=0}^n|\psi_i(t)\rangle\langle\psi_i(t)|\;.
\end{equation}
We can write
\begin{equation}
\hat{\rho}(t)=\sum_{n=0}^N d_n(t)\;\hat{\Pi}_n(t)\;,
\end{equation}
where
\begin{equation}
d_n(t)=p_n(t)-p_{n+1}(t)\geq0\;,
\end{equation}
so that
\begin{equation}
\frac{d}{dt}s_n(t)=\sum_{k=0}^N d_k(t)\;\mathrm{Tr}\left[\hat{\Pi}_n(t)\;\mathcal{L}\left(\hat{\Pi}_k(t)\right)\right]\;.
\end{equation}
With the explicit expression \eqref{lindblad} for $\mathcal{L}$, it is easy to prove that
\begin{equation}
\sum_{k=0}^N d_k(t)\;\mathrm{Tr}\left[\hat{\Pi}_n^\downarrow\;\mathcal{L}\left(\hat{\Pi}_k^\downarrow\right)\right]=(n+1)(s_{n+1}(t)-s_n(t))\;,
\end{equation}
so it would be sufficient to show that
\begin{equation}\label{PL}
\mathrm{Tr}\left[\hat{\Pi}_n(t)\;\mathcal{L}\left(\hat{\Pi}_k(t)\right)\right]\overset{?}{\leq} \mathrm{Tr}\left[\hat{\Pi}_n^\downarrow\;\mathcal{L}\left(\hat{\Pi}_k^\downarrow\right)\right]\;.
\end{equation}
We write explicitly the left-hand side of \eqref{PL}:
\begin{equation}\label{PLext}
\mathrm{Tr}\left[\hat{\Pi}_n(t)\;\hat{a}\;\hat{\Pi}_k(t)\;\hat{a}^\dag-\hat{\Pi}_n(t)\;\hat{\Pi}_k(t)\;\hat{a}^\dag\hat{a}\right]\;,
\end{equation}
where we have used that $\hat{\Pi}_n(t)$ and $\hat{\Pi}_k(t)$ commute.
\begin{itemize}
  \item Suppose $n\geq k$.
  Then
  \begin{equation}
  \hat{\Pi}_n(t)\;\hat{\Pi}_k(t)=\hat{\Pi}_k(t)\;.
  \end{equation}
  Using that
  \begin{equation}
  \hat{\Pi}_n(t)\leq\hat{\mathbb{I}}
  \end{equation}
  in the first term of \eqref{PLext}, we get
  \begin{equation}
  \mathrm{Tr}\left[\hat{\Pi}_n(t)\;\hat{a}\;\hat{\Pi}_k(t)\;\hat{a}^\dag-\hat{\Pi}_n(t)\;\hat{\Pi}_k(t)\;\hat{a}^\dag\hat{a}\right]\leq0\;.
  \end{equation}
  On the other hand, since the support of $\hat{a}\,\hat{\Pi}_k^\downarrow\,\hat{a}^\dag$ is contained in the support of $\hat{\Pi}_{k-1}^\downarrow$, and hence in the one of $\hat{\Pi}_n^\downarrow$, we have also
  \begin{equation}
  \hat{\Pi}_n^\downarrow\;\hat{a}\;\hat{\Pi}_k^\downarrow\;\hat{a}^\dag=\hat{a}\;\hat{\Pi}_k^\downarrow\;\hat{a}^\dag\;,
  \end{equation}
  so that
  \begin{equation}
  \mathrm{Tr}\left[\hat{\Pi}_n^\downarrow\;\hat{a}\;\hat{\Pi}_k^\downarrow\;\hat{a}^\dag-\hat{\Pi}_n^\downarrow\;\hat{\Pi}_k^\downarrow\;\hat{a}^\dag\hat{a}\right]=0\;.
  \end{equation}
  \item Suppose now that $k\geq n+1$.
  Then
  \begin{equation}
  \hat{\Pi}_n(t)\;\hat{\Pi}_k(t)=\hat{\Pi}_n(t)\;.
  \end{equation}
  Using that
  \begin{equation}
  \hat{\Pi}_k(t)\leq\hat{\mathbb{I}}
  \end{equation}
  in the first term of \eqref{PLext}, together with the commutation relation \eqref{CCR}, we get
  \begin{equation}
  \mathrm{Tr}\left[\hat{\Pi}_n(t)\;\hat{a}\;\hat{\Pi}_k(t)\;\hat{a}^\dag-\hat{\Pi}_n(t)\;\hat{\Pi}_k(t)\;\hat{a}^\dag\hat{a}\right]\leq n+1\;.
  \end{equation}
  On the other hand, since the support of $\hat{a}^\dag\,\hat{\Pi}_n^\downarrow\,\hat{a}$ is contained in the support of $\hat{\Pi}_{n+1}^\downarrow$ and hence in the one of $\hat{\Pi}_k^\downarrow$, we have also
  \begin{equation}
  \hat{\Pi}_k^\downarrow\;\hat{a}^\dag\;\hat{\Pi}_n^\downarrow\;\hat{a}=\hat{a}^\dag\;\hat{\Pi}_n^\downarrow\;\hat{a}\;,
  \end{equation}
  so that
  \begin{equation}
  \mathrm{Tr}\left[\hat{\Pi}_n^\downarrow\;\hat{a}\;\hat{\Pi}_k^\downarrow\;\hat{a}^\dag-\hat{\Pi}_n^\downarrow\;\hat{\Pi}_k^\downarrow\;\hat{a}^\dag\hat{a}\right]=n+1\;.
  \end{equation}
\end{itemize}
\subsection{Proof of Lemma \ref{lemma2}}
Since the quantum-limited attenuator is trace-preserving, we have
\begin{equation}
s_N(t)=\mathrm{Tr}\left[\hat{\rho}(t)\right]=1=s_N^\downarrow(t)\;.
\end{equation}
We will use induction on $n$ in the reverse order: suppose to have proved
\begin{equation}
s_{n+1}(t)\leq s_{n+1}^\downarrow(t)\;.
\end{equation}
We then have from \eqref{sdot}
\begin{equation}
\frac{d}{dt}s_n(t)\leq(n+1)\left(s_{n+1}^\downarrow(t)-s_n(t)\right)\;,
\end{equation}
while
\begin{equation}
\frac{d}{dt}s_n^\downarrow(t)=(n+1)\left(s_{n+1}^\downarrow(t)-s_n^\downarrow(t)\right)\;.
\end{equation}
Defining
\begin{equation}
f_n(t)=s_n^\downarrow(t)-s_n(t)\;,
\end{equation}
we have $f_n(0)=0$, and
\begin{equation}
\frac{d}{dt}f_n(t)\geq-(n+1)f_n(t)\;.
\end{equation}
This can be rewritten as
\begin{equation}
e^{-(n+1)t}\;\frac{d}{dt}\left(e^{(n+1)t}\;f_n(t)\right)\geq0\;,
\end{equation}
and implies
\begin{equation}
f_n(t)\geq0\;.
\end{equation}

\section{Generic one-mode Gaussian channels}\label{generic}
In this section we extend Theorem \ref{maintheorem} to any one-mode quantum Gaussian channel.

\begin{defn}
We say that two quantum channels $\Phi$ and $\Psi$ are equivalent if there are a unitary operator $\hat{U}$ and a unitary or anti-unitary $\hat{V}$ such that
\begin{equation}\label{Psi}
\Psi\left(\hat{X}\right)=\hat{V}\;\Phi\left(\hat{U}\;\hat{X}\;\hat{U}^\dag\right)\;\hat{V}^\dag
\end{equation}
for any trace-class operator $\hat{X}$.
\end{defn}
Clearly, a channel equivalent to a Fock-optimal channel is also Fock-optimal with a suitable redefinition of the Fock rearrangement:
\begin{lem}\label{U*}
Let $\Phi$ be a Fock-optimal quantum channel, and $\Psi$ be as in \eqref{Psi}.
Then, for any positive trace-class operator $\hat{X}$,
\begin{equation}
\Psi\left(\hat{X}\right)\prec_w\Psi\left(\hat{U}^\dag\left(\hat{U}\;\hat{X}\;\hat{U}^\dag\right)^\downarrow\hat{U}\right)\;.
\end{equation}
\end{lem}
The problem of analyzing any Gaussian quantum channel from the point of view of majorization is then reduced to the equivalence classes.

\subsection{Quadratures and squeezing}
In order to present such classes, we will need some more definitions.
The quadratures
\begin{eqnarray}
\hat{Q} &=& \frac{\hat{a}+\hat{a}^\dag}{\sqrt{2}}\\
\hat{P} &=& \frac{\hat{a}-\hat{a}^\dag}{i\sqrt{2}}
\end{eqnarray}
satisfy the canonical commutation relation
\begin{equation}
\left[\hat{Q},\;\hat{P}\right]=i\;\hat{\mathbb{I}}\;.
\end{equation}
In this section, and only in this section, $\hat{Q}$ and $\hat{P}$ will denote the above quadratures, and not generic projectors.
We can define a continuous basis of not normalizable vectors $\left\{|q\rangle\right\}_{q\in\mathbb{R}}$ with
\begin{eqnarray}
\hat{Q}|q\rangle &=& q|q\rangle\;,\\
\langle q|q'\rangle &=& \delta(q-q')\;,\\
\int_{\mathbb{R}}|q\rangle\langle q|\;dq &=& \hat{\mathbb{I}}\;,\\
e^{-iq\hat{P}}|q'\rangle &=& |q'+q\rangle\;,\qquad q,\,q'\in\mathbb{R}\;.
\end{eqnarray}
For any $\kappa>0$ we define the squeezing unitary operator \cite{barnett2002methods} $\hat{S}_\kappa$ with
\begin{equation}
\hat{S}_\kappa |q\rangle=\sqrt{\kappa}\;|\kappa q\rangle
\end{equation}
for any $q\in\mathbb{R}$.
It satisfies also
\begin{equation}
\hat{S}_\kappa^\dag\;\hat{P}\;\hat{S}_\kappa = \frac{1}{\kappa}\;\hat{P}\;.
\end{equation}

\subsection{Classification theorem}
Then, the following classification theorem holds \cite{holevo2007one,holevo2013quantum}:
\begin{thm}
Any one-mode quantum Gaussian channel is equivalent to one of the following:
\begin{enumerate}
\item a gauge-covariant Gaussian channel as in Definition \ref{gaugecovch} (cases $A_1)$, $B_2)$, $C)$ and $D)$ of \cite{holevo2007one});
\item a measure-reprepare channel $\Phi$ of the form
\begin{equation}\label{class2}
\Phi\left(\hat{X}\right)=\int_{\mathbb{R}}\langle q|\hat{X}|q\rangle\;e^{-iq\hat{P}}\;\hat{\rho}_0\;e^{iq\hat{P}}\;dq
\end{equation}
for any trace-class operator $\hat{X}$, where $\rho_0$ is a given Gaussian state (case $A_2)$ of \cite{holevo2007one});
\item a random unitary channel $\Phi_\sigma$ of the form
\begin{equation}\label{Phieta}
\Phi_\sigma\left(\hat{X}\right)=\int_{\mathbb{R}}e^{-iq\hat{P}}\;\hat{X}\;e^{iq\hat{P}}\;\frac{e^{-\frac{q^2}{2\sigma}}}{\sqrt{2\pi\sigma}}\;dq
\end{equation}
for any trace-class operator $\hat{X}$, with $\sigma>0$ (case $B_1)$ of \cite{holevo2007one}).
\end{enumerate}
\end{thm}
From Lemma \ref{U*}, with a suitable redefinition of Fock rearrangement all the channels of the first class are Fock-optimal.
On the contrary, for both the second and the third classes the optimal basis would be an infinitely squeezed version of the Fock basis:

\subsection{Class 2}
We will show that the channel \eqref{class2} does not have optimal inputs.

Let $\hat{\omega}$ be a generic quantum state.
Since $\Phi$ applies a random displacement to the state $\hat{\rho}_0$,
\begin{equation}\label{PhiX0}
\Phi\left(\hat{\omega}\right)\prec\hat{\rho}_0\;.
\end{equation}
Moreover, $\Phi\left(\hat{\omega}\right)$ and $\hat{\rho}_0$ cannot have the same spectrum unless the probability distribution $\langle q|\hat{\omega}|q\rangle$ is a Dirac delta, but this is never the case for any quantum state $\hat{\omega}$, so the majorization in \eqref{PhiX0} is always strict.
Besides, in the limit of infinite squeezing the output tends to $\hat{\rho}_0$ in trace norm:
\begin{align}
&\left\|\Phi\left(\hat{S}_\kappa\;\hat{\omega}\;\hat{S}_\kappa^\dag\right)-\hat{\rho}_0\right\|_1\nonumber\\
&=\left\|\int_{\mathbb{R}}\langle q|\hat{\omega}|q\rangle\left(e^{-i\kappa q\hat{P}}\;\hat{\rho}_0\;e^{i\kappa q\hat{P}}-\hat{\rho}_0\right)dq\right\|_1\nonumber\\
&\leq\int_{\mathbb{R}}\langle q|\hat{\omega}|q\rangle\left\|e^{-i\kappa q\hat{P}}\;\hat{\rho}_0\;e^{i\kappa q\hat{P}}-\hat{\rho}_0\right\|_1dq\;,
\end{align}
and the last integral tends to zero for $\kappa\to0$ since the integrand is dominated by the integrable function $2\langle q|\hat{\omega}|q\rangle$, and tends to zero pointwise.
It follows that the majorization relation
\begin{equation}
\Phi\left(\hat{S}_\kappa\;\hat{\omega}\;\hat{S}_\kappa\right)\prec\Phi\left(\hat{\omega}\right)
\end{equation}
will surely not hold for some positive $\kappa$ in a neighbourhood of $0$, and $\hat{\omega}$ is not an optimal input for $\Phi$.

\subsection{Class 3}
For the channel \eqref{Phieta}, squeezing the input always makes the output strictly less noisy.
Indeed, it is easy to show that for any positive $\sigma$ and $\sigma'$
\begin{equation}
\Phi_\sigma\circ\Phi_{\sigma'}=\Phi_{\sigma+\sigma'}\;.
\end{equation}
Then, for any $\kappa>1$ and any positive trace-class $\hat{X}$
\begin{align}
\hat{S}_\kappa\;\Phi_\sigma\left(\hat{X}\right)\;\hat{S}_\kappa^\dag &= \Phi_{\kappa^2\sigma}\left(\hat{S}_\kappa\;\hat{X}\;\hat{S}_\kappa^\dag\right)\nonumber\\
&=\Phi_{(\kappa^2-1)\sigma}\left(\Phi_{\sigma}\left(\hat{S}_\kappa\;\hat{X}\;\hat{S}_\kappa^\dag\right)\right)\;,
\end{align}
hence, recalling that $\Phi$ applies a random displacement,
\begin{equation}
\Phi_\sigma\left(\hat{X}\right)\prec \Phi_{\sigma}\left(\hat{S}_\kappa\;\hat{X}\;\hat{S}_\kappa^\dag\right)\;.
\end{equation}

\section{The thinning}\label{secthinning}
The thinning \cite{renyi1956characterization} is the map acting on classical probability distributions on the set of natural numbers that is the discrete analogue of the continuous rescaling operation on positive real numbers.

In this Section we show that the thinning coincides with the restriction of the Gaussian quantum-limited attenuator to quantum states diagonal in the Fock basis, and we hence extend Theorem \ref{maintheorem} to the discrete classical setting.

\begin{defn}[$\ell^1$ norm]
The $\ell^1$ norm of a sequence $\{x_n\}_{n\in\mathbb{N}}$ is
\begin{equation}
\|x\|_1=\sum_{n=0}^\infty |x_n|\;.
\end{equation}
We say that $x$ is summable if $\|x\|_1<\infty$.
\end{defn}
\begin{defn}
A discrete classical channel is a linear positive map on summable sequences that is continuous in the $\ell^1$ norm and preserves the sum, i.e. for any summable sequence $x$
\begin{equation}
\sum_{n=0}^\infty\left[\Phi(x)\right]_n=\sum_{n=0}^\infty x_n\;.
\end{equation}
\end{defn}
The definitions of passive-preserving and Fock-optimal channels can be easily extended to the discrete classical case:
\begin{defn}
Given a summable sequence of positive numbers $\{x_n\}_{n\in\mathbb{N}}$, we denote with $x^\downarrow$ its decreasing rearrangement.
\end{defn}
\begin{defn}
We say that a discrete classical channel $\Phi$ is passive-preserving if for any decreasing summable sequence $x$ of positive numbers $\Phi(x)$ is still decreasing.
\end{defn}
\begin{defn}
We say that a discrete classical channel $\Phi$ is Fock-optimal if for any summable sequence $x$ of positive numbers
\begin{equation}\label{optimalcl}
\Phi(x)\prec\Phi\left(x^\downarrow\right)\;.
\end{equation}
\end{defn}
Let us now introduce the thinning.
\begin{defn}[Thinning]
Let $N$ be a random variable with values in $\mathbb{N}$.
The thinning with parameter $0\leq\lambda\leq1$ is defined as
\begin{equation}
T_\lambda(N)=\sum_{i=1}^N B_i\;,
\end{equation}
where the $\{B_n\}_{n\in\mathbb{N}^+}$ are independent Bernoulli variables with parameter $\lambda$, i.e. each $B_i$ is one with probability $\lambda$, and zero with probability $1-\lambda$.
\end{defn}
From a physical point of view, the thinning can be understood as follows:
consider a beam-splitter of transmissivity $\lambda$, where each incoming photon has probability $\lambda$ of being transmitted, and $1-\lambda$ of being reflected, and suppose that what happens to a photon is independent from what happens to the other ones.
Let $N$ be the random variable associated to the number of incoming photons, and $\{p_n\}_{n\in\mathbb{N}}$ its probability distribution, i.e. $p_n$ is the probability that $N=n$ (i.e. that $n$ photons are sent).
Then, $T_\lambda(p)$ is the probability distribution of the number of transmitted photons.
It is easy to show that
\begin{equation}\label{Tn}
\left[T_\lambda(p)\right]_n=\sum_{k=0}^\infty r_{n|k}\;p_k\;,
\end{equation}
where the transition probabilities $r_{n|k}$ are given by
\begin{equation}\label{rnk}
r_{n|k}=\binom{k}{n}\lambda^n(1-\lambda)^{k-n}\;,
\end{equation}
and vanish for $k<n$.

The map \eqref{Tn} can be uniquely extended by linearity to the set of summable sequences:
\begin{equation}\label{Tne}
\left[T_\lambda(x)\right]_n=\sum_{k=0}^\infty r_{n|k}\;x_k\;,\qquad \|x\|_1<\infty\;.
\end{equation}
\begin{prop}
The map $T_\lambda$ defined in \eqref{Tne} is continuous in the $\ell^1$ norm and sum-preserving.
\begin{proof}
For any summable sequence $x$ we have
\begin{equation}
\sum_{n=0}^\infty\left|T_\lambda(x)\right|_n\leq\sum_{n=0}^\infty\sum_{k=0}^\infty r_{n|k}\;|x_k|=\sum_{k=0}^\infty|x_k|\;,
\end{equation}
where we have used that for any $k\in\mathbb{N}$
\begin{equation}
\sum_{n=0}^\infty r_{n|k}=1\;.
\end{equation}
Then, $T_\lambda$ is continuous in the $\ell^1$ norm.

An analogous proof shows that $T_\lambda$ is sum-preserving.
\end{proof}
\end{prop}

\begin{thm}\label{thinatt}
Let $\Phi_\lambda$ and $T_\lambda$ be the quantum-limited attenuator and the thinning of parameter $0\leq\lambda\leq1$, respectively.
Then for any summable sequence $x$
\begin{equation}
\Phi_\lambda\left(\sum_{n=0}^\infty x_n\;|n\rangle\langle n|\right)=\sum_{n=0}^\infty \left[T_\lambda(x)\right]_n\;|n\rangle\langle n|\;.
\end{equation}
\begin{proof}
Easily follows from the representation \eqref{kraus}, \eqref{Tn} and \eqref{rnk}.
\end{proof}
\end{thm}
As easy consequence of Theorem \ref{thinatt} and Theorem \ref{maintheorem}, we have
\begin{thm}
The thinning is passive-preserving and Fock-optimal.
\end{thm}

\section{Conclusions}
We have proved that for any one-mode gauge-covariant bosonic Gaussian channel, the output generated by any state diagonal in the Fock basis and with decreasing eigenvalues majorizes the output generated by any other input state with the same spectrum.
Then, the input state with a given entropy minimizing the output entropy is certainly diagonal in the Fock basis and has decreasing eigenvalues.
The non-commutative quantum constrained minimum output entropy problem is hence reduced to a problem in classical discrete probability, that for the quantum-limited attenuator involves the thinning channel, and whose proof could exploit the techniques of the proof of the Restricted Thinned Entropy Power Inequality \cite{johnson2010monotonicity}.

Exploiting unitary equivalence we also extend our results to one-mode trace-preserving bosonic Gaussian channel which are not gauge-covariant, with the notable exceptions of those special maps admitting normal forms $A_2)$ and $B_1)$ \cite{holevo2007one} for which we show that no general majorization ordering is possible.

\section*{Acknowledgment}
The Authors  thank Andrea Mari, Luigi Ambrosio, Seth Lloyd and Alexander S. Holevo for comments and fruitful discussions.
GdP thanks G. Toscani and G. Savar\'e for the ospitality and the useful discussions in Pavia.

\bibliographystyle{IEEEtran}
\bibliography{biblio}

\begin{IEEEbiography}[{\includegraphics[width=1in,height=1.25in,clip,keepaspectratio]{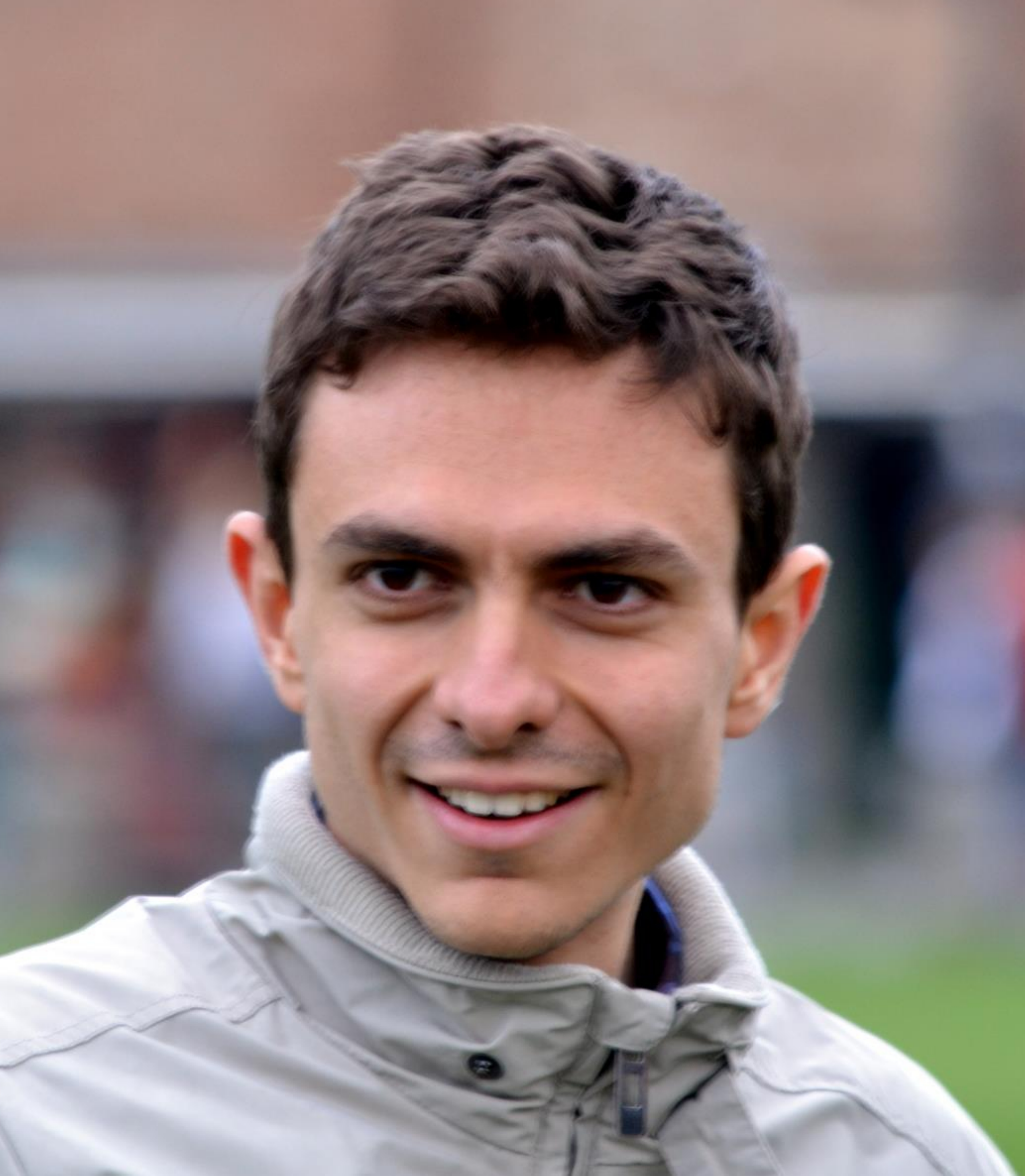}}]{Giacomo De Palma}
was born in Lanciano (CH), Italy, on March 15, 1990.
He received the B.S. degree in Physics and the M.S. degree in Physics from the University of Pisa, Pisa (PI), Italy, in 2011 and 2013, respectively. He also received the ``Diploma di Licenza'' in Physics from Scuola Normale Superiore, Pisa (PI), Italy, in 2014.

He is getting the Ph.D. degree in Physics at Scuola Normale Superiore in 2016.
His research interests include quantum information, quantum statistical mechanics and quantum thermodynamics.
He is author of nine scientific papers published in peer-reviewed journals.
\end{IEEEbiography}
\begin{IEEEbiography}[{\includegraphics[width=1in,height=1.25in,clip,keepaspectratio]{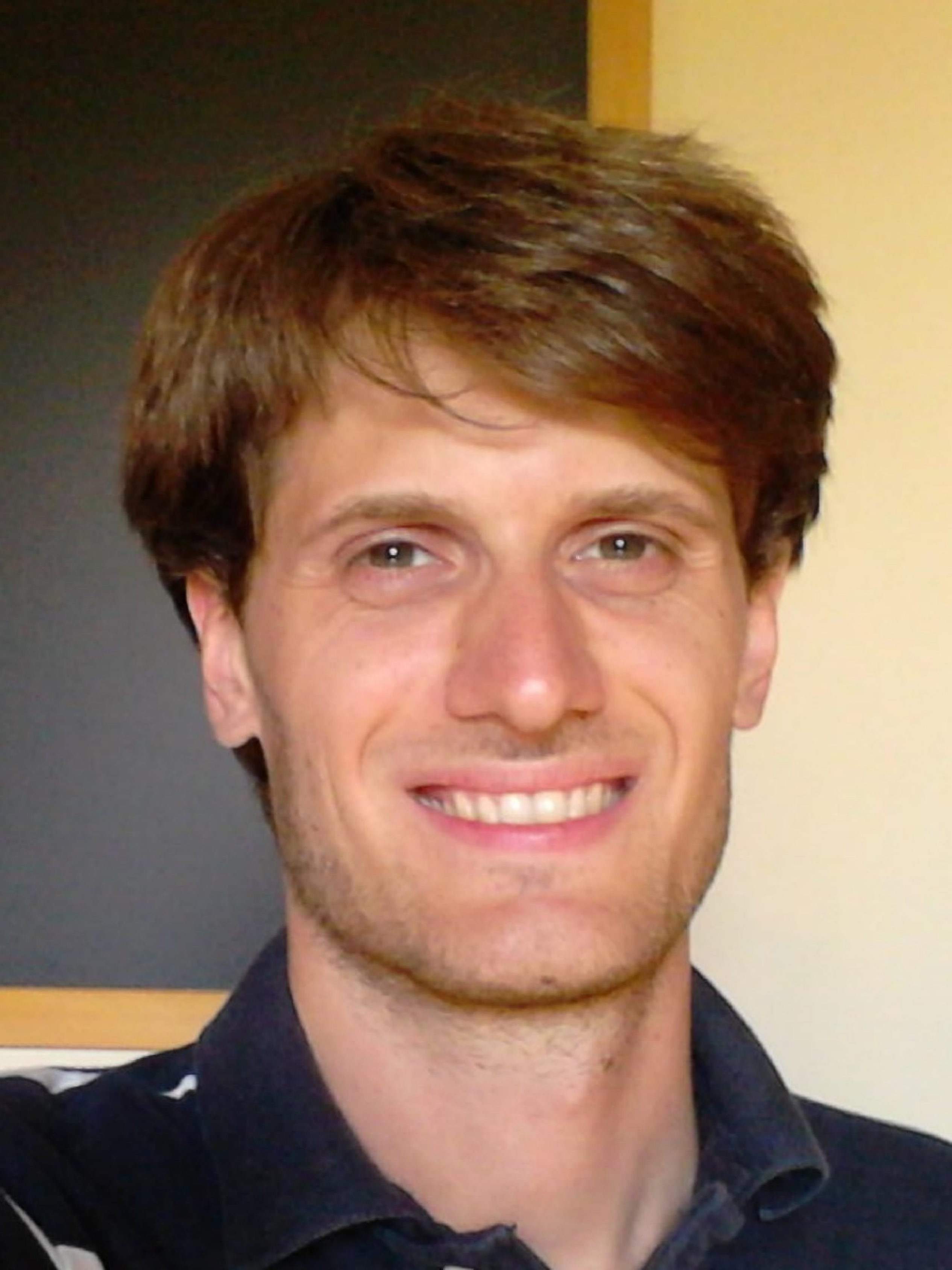}}]{Dario Trevisan}
was born in the Province of Venice, Italy. He received the B.S. degree in mathematics from the University of Pisa, Pisa, Italy, in 2009, the M.S. degree in mathematics from the University of Pisa, in 2011, and the Ph.D. degree in mathematics from the Scuola Normale Superiore, Pisa, Italy, in 2014.

He is currently Assistant Professor at the University of Pisa.

Dr. Trevisan is a member of the GNAMPA group of the Italian National Institute for Higher Mathematics (INdAM).
\end{IEEEbiography}
\begin{IEEEbiography}[{\includegraphics[width=1in,height=1.25in,clip,keepaspectratio]{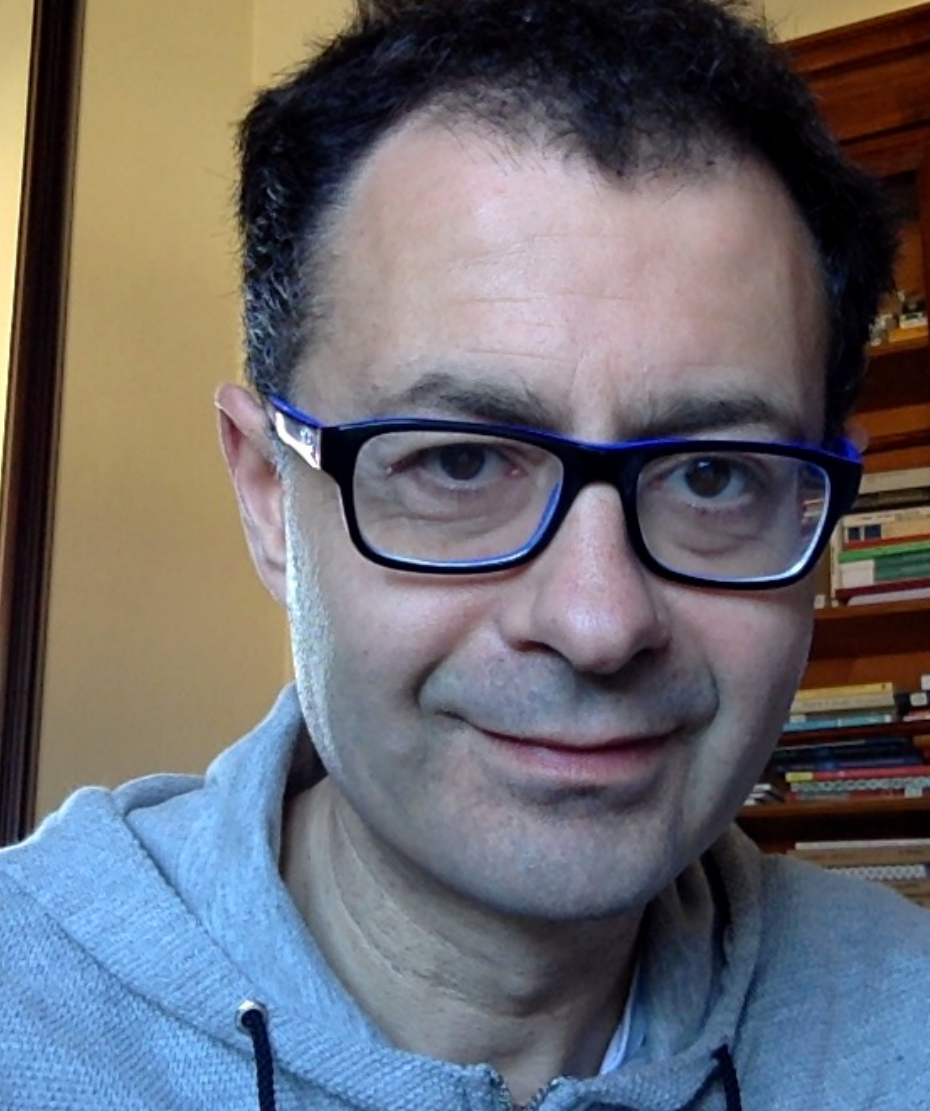}}]{Vittorio Giovannetti}
was born in Castelnuovo di Garfagnana (LU) Italy, on April 1, 1970. He received the M.S. degree in Physics from the University of Pisa and PhD degree in theoretical Physics from the University of Perugia.

He is currently Associate Professor at the Scuola Normale Superiore of Pisa.
\end{IEEEbiography}
\end{document}